\documentclass[aps,prd,preprintnumbers,showpacs]{revtex4}
\usepackage[dvips]{graphicx}
\begin{document}

%
%  Uncomment following two lines and one below for 2 column format.
%@twocolumnfalse\endcsname

\eprint{Nisho-2-2015}
\title{FRBs and dark matter axions }
%\author{Hirotsugu, Fujii}
%\address{Institute of Physics, University of Tokyo, Komaba, Meguro-ku, Tokyo, 153-8902,  Japan}
%\author{Kazunori, Itakura}
%\address{Institute of Particle and  Nuclear Studies, High Energy
%Accelerator Research Organization, 1-1, Ooho, Tsukuba, Ibaraki,
%305-0801, Japan}
\author{Aiichi Iwazaki}
\affiliation{International Economics and Politics, Nishogakusha University,\\ 
6-16 3-bantyo Chiyoda Tokyo 102-8336, Japan.}   
\date{Dec. 19, 2015}
\begin{abstract}
We have proposed a model of a progenitor of fast radio bursts (FRBs):
The FRBs are emitted by electrons in atmospheres of neutron stars
when neutron stars collide with dark matter axion stars. 
We reexamine the model by taking account of the tidal forces of the neutron stars.
The axion stars are distorted by the forces so that they become like long sticks 
when they are close to the neutron stars.
Although the tidal forces fairly distort the axion stars, our model is still 
consistent with present observations. We explain a distinctive feature of our model that
the FRBs are not broadband, but narrowband. 
We also show that two-component FRBs may arise when the axion stars collide with
binary neutron stars. FRBs would most frequently arise in centers of galaxies 
since the dark matter concentrates in the centers.
\end{abstract}
\hspace*{0.3cm}

\hspace*{1cm}

\maketitle

%\vskip2pc
%%%%%%%%%%%%%%%%%%%%%%%%%%%%%%%%%%%%%%%%%%%%%%%
Fast Radio Bursts have recently been discovered\cite{frb,frb1,frb2} at around $1.4$ GHz frequency.
The durations of the bursts are typically
a few milliseconds. The origin of the bursts has been suggested to be
extra-galactic owing to their large dispersion measures.
This suggests that the large amount of the energies $\sim 10^{40}$erg
are produced at the radio frequencies.
The event rate of the burst has been estimated to be $\sim 10^{-3}$ per year in a galaxy.
Furthermore, no gamma or X ray bursts associated with the bursts
have been detected. Follow up observations\cite{frb3} of FRBs do not find any 
signals from the directions of the FRBs. 
To find progenitors of the bursts, several models\cite{model} have been proposed.
They ascribe FRBs to traditional sources such as neutron star-neutron star mergers, 
magnetors, black holes, et al.. 

Our model\cite{i,t} proposed recently ascribes FRBs to axions\cite{axion}, 
which are one of most promising candidates for
dark matter. A prominent feature of axions is that they are converted to 
radiations under strong magnetic fields. 
According to the model,
the FRBs arise from the collisions between axion stars and neutron stars.  
They are emitted by electrons accelerated by electric fields induced on the axion stars.
We have explained the observed properties such as
duration, event rate and total radiation energy of the FRBs.
But, we have neglected the effects of the tidal forces of the neutron stars
in the model.

\vspace{0.1cm}
In this paper we reexamine the model by taking account of the tidal forces.
Although the tidal forces distort the axion stars to look
like long sticks, we show that the model is still consistent with the observations.
It apparently seems that FRBs are monochromatic in our model 
because electrons emitting the FRBs oscillate with
the frequency $m_a/2\pi$ given by the axion mass $m_a$.
We explain in our model that 
the FRBs have finite bandwidth owing to the collisional effects of the electrons.
The FRBs are redshifted gravitationally and cosmologically so that the frequencies detected at the earth are
smaller than $m_a/2\pi\simeq 2.4$GHz ( $m_a/10^{-5}\rm eV$ ).
The collisions generating the FRBs would most frequently arise in centers of galaxies because the dark matter
concentrates in the centers.
Furthermore, we point out that a two-component FRB recently observed\cite{two} may arise
from the collision between an axion star and a binary neutron star.

\vspace{0.2cm}
We first make a brief review of the our original model without the effects of the tidal forces.
Then, we show how the effects of the tidal forces modify the original model.
We first start to make a review of axions and axion stars.

\vspace{0.1cm}
Axions described by a real scalar field $a(\vec{x},t)$ are Nambu-Goldstone boson 
associated with Pecci-Quinn global U(1) symmetry\cite{PQ}.
The symmetry was introduced to cure strong CP problems in QCD.
The axions are massless when they are born in the early Universe.
Then, the potential term $-f_a^2m_a^2\cos(a/f_a)$
develops owing to the interaction with instantons in QCD at the temperature below $1$GeV, where
$f_a$ denotes the decay constant of the axions. 
Thus, the axion field oscillates around the minimum $a=0$ of the potential.
It can take
a different initial value in a region from those in other regions causally disconnected.  
Thus, there are many regions causally disconnected 
at the epoch below the temperature $1$GeV, 
in each of which the axion field takes
a different oscillation phase; energy density is also
different.
With the expansion of the Universe, the regions with different energy densities
are causally connected. Thus, there arises spatial fluctuations of the axion energy density.
The nonlinear effects of the axion potential make the fluctuations with
over densities in some regions
grow to form axion miniclusters\cite{kolb} at
the period of equal axion-matter radiation energy density in the regions.
Their masses have been estimated to be of the order of $10^{-12}M_{\odot}$.
Furthermore, these miniclusters condense to form axion stars
by gravitationally losing their kinetic energies\cite{axions}.
Therefore, the masses of axion stars are expected to be of the order of $10^{-12}M_{\odot}$.

\vspace{0.1cm}
The axion stars are gravitationally bounded states of the coherent axions.
They are
known as oscillaton\cite{axion2} and
are described as 
classical solutions\cite{i,axions,osc,iwa,osci} of the real scalor field $a(\vec{x},t)$ coupled with gravity.
In particular, solutions describing spherical symmetric axion stars with small masses 
are given by\cite{i} 

\begin{equation}
\label{a2}
a(\vec{x},t)=a_0f_a\exp(-\frac{r}{R_a})\cos(m_at),
\end{equation} 
where $R_a$ denotes radii of the axion stars, which are related with
the masses $M_a$ of the axion stars,

\begin{equation}
\label{R}
R_a=\frac{m^2_{\rm pl}}{m_a^2M_a}
\simeq 260\,\mbox{km}\,\,\Big(\frac{10^{-5}\rm eV}{m_a}\Big)^2\Big(\frac{10^{-12}M_{\odot}}{M_a}\Big). 
\end{equation} 

The amplitude $a_0$ is given by

\begin{equation}
\label{a_0}
a_0\simeq 0.9\times 10^{-6} \Big(\frac{10^2\,\mbox{km}}{R_a}\Big)^2\frac{10^{-5}\mbox{eV}}{m_a}.
\end{equation}

The solutions are obtained\cite{i} 
by neglecting terms of higher orders of $a/f_a$ in the potential term 
$-f_a^2m_a^2\cos(a/f_a)=-f_a^2m_a^2+m_a^2a^2/2+``\mbox{higher orders}"$ except for the mass term $m_a^2a^2/2$.
Actually,
the condition $a/f_a\ll 1$ is satisfied 
for the small mass $M_a \sim 10^{-12}M_{\odot}$.
These axion stars with small masses are gravitationally loosely bounded states of axions.

\vspace{0.1cm}
We should note that the axion field representing the axion stars oscillates with the frequency $m_a/2\pi$.
Then, it follows that the axion field generates oscillating electric fields $E_a$
under background magnetic fields $B$,

\begin{eqnarray}
\label{ele}
\vec{E}_a(r,t)&=&-\alpha \frac{a(\vec{x},t)\vec{B}}{f_a\pi}
=-\alpha \frac{a_0\exp(-r/R_a)\cos(m_at)\vec{B}(\vec{r})}{\pi}\\
&\simeq& 0.4\times 1\mbox{eV}^2( \,\,=2\times 10^4\mbox{eV}/\mbox{cm} \,\,)\cos(m_at)
\Big(\frac{10^2\,\mbox{km}}{R_a}\Big)^2\frac{10^{-5}\mbox{eV}}{m_a}\frac{B}{10^{10}\mbox{G}},
\end{eqnarray} 
where the background magnetic fields are supposed to be given by neutron stars.

The generation is caused by the interaction between axions and electromagnetic gauge fields,

\begin{equation}
\label{L}
L_{aEB}=k\alpha \frac{a(\vec{x},t)\vec{E}\cdot\vec{B}}{f_a\pi}
\end{equation}
with the fine structure constant $\alpha\simeq 1/137$,   
where the numerical constant $k$ depends on axion models; typically it is of the order of one.
Hereafter we set $k=1$.
This oscillating electric fields make electrons emit coherent radiations when the axion stars collide with neutron stars.

It should be mentioned that the number of the axions in the volume $\lambda^3\equiv (2\pi/m_a)^3$
is huge such as $10^{39}(10^{-5}\mbox{eV}/m_a)^{6}$. It implies that the axions forming
the axion stars are coherent. Thus, such axions can be described as the classical field $a(\vec{x},t)$.
These coherent axions are converted 
to the electric fields under the background magnetic field.

\vspace{0.1cm}
We use the value $M_a \sim 10^{-12}M_{\odot}$ as a reference of the masses of the axion stars.
The value has been obtained by the comparison between theoretical and observational event rates of FRBs
in our model\cite{i}, although
we have derived it by neglecting the tidal forces of the neutron stars. It is remarkable that
the mass is coincident with the mass of axion miniclusters obtained previously.
% (As we show below, even if the tidal forces are taken into account, the value $M_a \sim 10^{-12}M_{\odot}$
% is still reasonable. Similarly, we may use the value $a_0$ given above even if
% the axion stars are distorted by the tidal forces. )
Here, we breifly show how we obtain the masses of the axion stars in the original model.
Later we show that the result does not change even if we take into account the tidal forces.

In order to obtain the masses, 
we derive the event rate $R_{\rm burst}$ of the collisions between axion stars and neutron stars.
Assuming that halo of a galaxy is composed of axion stars and that local density
of the halo is given by $0.5\times 10^{-24}\,\mbox{g\,cm}^{-3}$,
the event rate of the collisions in a galaxy per year is given by

\begin{equation}
R_{\rm burst}=n_a\times N_{\rm ns}\times Sv\times 1\rm year,
\end{equation}
with relative velocity $v=300\rm km/s$ between axion stars and neutron stars,
where $n_a$( $=0.5\times 10^{-24}\,\mbox{g\,cm}^{-3}/M_a$ ) denotes the number density of the axion stars and 
$N_{\rm ns}$ represents the number of neutron stars in a galaxy;
it is supposed to be $10^9$.
The cross section $S$ for the collision is given by
$S=\pi (R_a+R_{ns})^2\big(1+2GM_{ns}/v^2(R_a+R_{ns})\big)\simeq 2.8\pi (R_a+R_{ns})GM_{\odot}/v^2$ 
where $R_{ns}\,(=10$km ) denotes the radius of neutron star
with mass $M_{ns}=1.4M_{\odot}$. This cross section is obtained by neglecting
the effects of the tidal forces.
Using the formula, we
find the mass $\sim 10^{-12}M_{\odot}$ of the axion stars,
by comparing the observed rate ( $10^{-3}$ per galaxy and per year ) of the FRBs with
the theoretical formula $R_{\rm burst}$,

\begin{eqnarray}
R_{\rm burst}&=&\frac{0.5\times 10^{-24}\,\mbox{g\,cm}^{-3}}{M_a}\times 10^9\times 
2.8\pi(10\mbox{km}+R_a)\frac{GM_{\odot}}{10^{-6}}\times 1\rm year \nonumber \\ 
&\sim& 
10^{-3}\Big(\frac{10^{-12}M_{\odot}}{M_a}\Big)\frac{
10\mbox{km}+260\mbox{km}\Big(\frac{10^{-5}\mbox{eV}}{m_a}\Big)^2\frac{10^{-12}M_{\odot}}{M_a}}
{10\mbox{km}+260\mbox{km}}.
\label{Rb}
\end{eqnarray}  

The parameters used above involve several ambiguities. 
For example, both of local halo density and number density of neutron stars in a galaxy 
are not well known.  Furthermore, 
the event rate of the FRBs ( $\sim 10^{-3}$ in a year per a galaxy ) has not yet been correctly determined. 
Thus,
the observed rate only constrains the masses of the axion stars in a range such that 
$M_a=10^{-13}M_{\odot}\sim 10^{-11}M_{\odot}$. As we show below,
these ambiguities allow us to determine the mass of the axion stars such as $M_a\sim 10^{-12}M_{\odot}$
even if we take account of the tidal forces of the neutron stars.

\vspace{0.1cm}
Now we show how large amount of energies are emitted by the collision between the axion stars and the neutron stars.
Because the atmospheres of the neutron stars involve dense electrons,
the coherent radiations are
emitted by electrons in the atmospheres with strong magnetic fields $B\sim 10^{10}$G. 
These electrons are coherently oscillated 
by the electric fields $E_a$ induced on the axion stars under the magnetic fields $B$.
Each electron oscillating with the frequency $m_a/2\pi$ emits a dipole radiation
with the rate,

\begin{equation}
\dot{w}\equiv \frac{2e^2\dot{p}^2}{3 m_e^2}=\frac{2e^2(e\alpha a_0 B/\pi)^2}{3m_e^2}
\simeq 0.7\times 10^{-9}\mbox{GeV/s}\Big(\frac{10^2\rm km}{R_a}\Big)^4\Big(\frac{10^{-5}\rm eV}{m_a}\Big)^2
\Big(\frac{B}{10^{10}\mbox{G}}\Big)^2
\end{equation}
with electron mass $m_e$,
where $p$ denotes the momentum of the electron parallel to $\vec{E}_a$ governed by the
equation of motion, $dp/dt=\dot{p}=-eE_a$.
The magnetic fields of the neutron stars does not affect the equation because they are parallel to $\vec{E}$. 
Because the electrons in the volume $\lambda^3\equiv (2\pi/m_a)^3$ coherently oscillate, 
the total emission rate $\dot{W}$ from the electrons is given such that
 $\dot{W}=\dot{w}(n_e\lambda^3)^2=2(n_e\lambda^3)^2\dot{p}^2/(3 m_e^2)$,
where $n_e$ denotes the number density of electrons.  In other words,
the emission rate $\dot{W}/\lambda^3$ per unit volume is given by,

\begin{equation}
\label{w}
\dot{W}/\lambda^3 
\sim 10^{34}\,\mbox{GeV}/\mbox{(s cm$^3$)}\,\Big(\frac{n_e}{10^{20}\mbox{cm}^{-3}}\Big)^2
\,\Big(\frac{10^{-5}\rm eV}{m_a}\Big)^6\Big(\frac{B}{10^{10}\mbox{G}}\Big)^2.
\end{equation} 
The value should be compared with the energy densities $\rho_a$ of the axion stars; 
$\rho_a\sim M_a/R_a^3\simeq 10^{25}\rm GeV/cm^3$.
When the axion stars collide with the neutron stars, the relative velocities are
given by $v_s=\sqrt{2G(1.4M_{\odot})/R_{ns}}\simeq 6\times 10^{-1}\simeq 2\times 10^{10}$cm/s,
where the mass ( radius ) of the neutron stars is supposed to be
$1.4M_{\odot}$ ( $10$km ), respectively.
It takes $10^{-10}$s for them to pass the distance $1$cm.
Thus, the axion stars evaporate in the depth 
with approximately $n_e\sim 10^{20}\rm cm^{-3}$, when they collide with the surfaces of
the neutron stars.

Because the relative velocities is 
about $10^5$km/s, it takes a millisecond for the neutron stars to 
pass the axion stars with radus $10^2$km. The emission lasts for the period.
Furthermore, the total amount energies emitted in the collisions 
are given by $M_a(10\rm km/10^2km)^2\sim 10^{40}erg$.
This is our previous result obtained by neglecting the tidal forces of the neutron stars.
Obviously, 
the axion stars are extremely deformed by the tidal forces so that 
the results of the duration and the total amount of energies
are not reliable.

\vspace{0.2cm}
Up to now, we have made a sketch of our original model of the FRBs
without including the effects of the tidal forces.
As we can see below, almost all of the results shown above hold even if the tidal forces are important.

Obviously, when the axion stars are sufficiently far away from the neutron stars,
the tidal forces of the neutron stars are much weaker than the
self-gravity. On the other hand, when they approach the neutron stars, the tidal forces
are important.
We first derive the distances of axion stars from neutron stars at which 
the tidal forces of the neutron stars are stronger than the self-gravity of the axion stars.
Attractive forces per unit mass at the surface of the axion stars by the axion stars themselves
are given by $GM_a/R_a^2$. On the other hand, the difference between the gravitational
forces at the surface and the ones at the center of the axion stars exerted by the neutron stars
at the distance $r_c (\,\,\gg R_a \,\,)$ from the axion stars is given by

\begin{equation}
\frac{GM_{\rm ns}}{r_c^2} -\frac{GM_{\rm ns}}{(r_c+R_a)^2}
\simeq \frac{GM_{\rm ns}}{r_c^2}\frac{2R_a}{r_c}
\end{equation}
Thus, when the axion stars approach 
the neutron stars within the distance $r_c=R_a(2M_{\rm ns}/M_a)^{1/3}$,
the tidal forces of the neutron stars are 
stronger than the self-gravity. Then, 
the axions forming the axion stars freely fall toward the neutron stars. 
Numerically $r_c$ is equal to $1.4\times 10^6$km when $M_{\rm ns}=1.4M_{\odot}$ and
$R_a=10^2$km.

\vspace{0.1cm}
In order to see the event rate of the collisions and find the mass of the axion stars,
we need to derive the scattering cross section $S_{\rm re}$ between the axion stars and the neutron stars
when the tidal forces are important. It turns out that
the cross section is $10$ times smaller than the one shown above.

For the purpose, we would like to
obtain a scattering cross section $S_c$ of
the axion stars at the distance $r_c$
from the neutron stars. From the location, 
the axions freely fall toward the neutron stars.
We suppose that the cross section $S_c$ is represented by $\pi L_c^2$.
Then, from the conservation of the angular momentum, $v_cL_c=v_sR_{\rm ns}$
and the energy conservation $v_c^2/2-GM_{\rm ns}/r_c=v_s^2/2-GM_{\rm ns}/R_{\rm ns}$, 
we find that $L_c=R_{\rm ns}\sqrt{1+2GM_{\rm ns}/(R_{\rm ns}v_c^2)}$,
where $v_c$ denotes the velocity of the axion stars at the location and
$v_s$ does the velocity at the surface of the neutron stars.    
Thus, we obtain $S_c=2\pi GM_{\rm ns}R_{\rm ns}/v_c^2$ where we use the inequality 
$1\ll GM_{\rm ns}/(R_{\rm ns}v_c^2)$. 
It should be compared with the cross section $S=2\pi GM_{\rm ns}R_a/v^2$ 
shown above; $S_c/S=(v^2/v_c^2)R_{\rm ns}/R_a$.
( Although we used the non-relativistic energies, the result is essencially 
identical to the one obtained by using relativistic ones. )

Next we calculate the real cross section 
$S_{\rm re}=\pi L^2$
between the axion stars and the neutron stars separated infinitely with each other, where
$L$ denotes the impact parameter in the collision.
Using the conservation of the angular momentum $v_cL_c=vL$,
we obtain $S_c/S_{\rm re}=(v/v_c)^2$ where $v_c\simeq 600$km/s since we assume $v=300$km/s. 
Thus,
the real cross section is given by $S_{\rm re}=SR_{\rm ns}/R_a\simeq S/10$. 
That is, the cross section under the influence of the tidal forces is $10$ times smaller 
than the one under no influence of the tidal forces.
It follows that 
the event rate of the FRBs becomes
10 times smaller than the rate found in eq(\ref{Rb}).
Because there are several ambiguities in the parameters used to
estimate the masses of the axion stars, 
the effects of the tidal forces
do not seriously change the estimation.
Therefore, even if we take into account the effects of the tidal forces,
it is reasonable that we estimate the masses of the axion stars such as
$M_a\sim 10^{-12}M_{\odot}$. 

\vspace{0.1cm}
We proceed to show how the axion stars are distorted by the tidal forces and collide with
the neutron stars. For simplicity,
we consider the collision with vanishing impact parameter, that is,
the central collision.
At the distance $r_c\simeq 10^6$km from the neutron stars, 
the axion stars are spherical with the radius $R_a=10^2$km 
and their velocities are given by $v_c\simeq 600$km/s.
Although they keep spherical forms up to the distance, they can not keep it
at the distances less than $r_c$. 
Obviously the center of the axion stars falls straight toward the center of the neutron stars.
On the other hand, the edge of the axion stars with its impact parameter $R_a$
falls in a location on their surfaces with the velocity $v_s'$, 
whose location on the surfaces
is at the distance $r_n$ measured 
from the line connecting both centers of the axion stars and neutron stars.
Then, from the conservation of the angular momentum $v_s'r_n=v_cR_a$ and 
the energy conservation, we obtain $r_n\simeq 0.36$km and $v_s'\simeq 1.6\times 10^5$km/s.
Therefore, it turns out that the axion stars become like long sticks approximately 
with dimensions $(0.5\rm km)^2\times 10^5$km when they collide with the neutron stars.

Here, the length $\sim 10^5$km of the stretched axion stars is 
determined in the following.
Because it approximately takes $0.3$ second for the collision to be accomplished
from the beginning of the collision through the end of the collision,
the length is given by $0.3\rm s\times 1.6\times 10^5km/s\sim 10^5$km.
The time $0.3$s is approximately equal to the time 
it takes for the axion stars located at the distance $r_c$ to pass the distance $2R_a=200$km 
with the velocity $v_c=600$km/s.

\vspace{0.1cm} 
Although the form of the axion stars is stretched by the tidal forces,
the coherence of the axions is still kept since the number of the axions involved
in the volume $(2\pi/m_a)^3$ is still extremely large. Thus, we may use the classical field $a(\vec{x},t)$
in order to represent the axion stars. 

\vspace{0.1cm}
Now, we would like to see 
the emission rate of the radiations from the stretched axion stars when they collide with the neutron stars.
For the purpose, we need to know the strength of the electric fields induced on the axion stars.
The electric field induced on the spherical axion stars is given in eq(\ref{ele}).
On the other hand, the electric field induced on the stretched axion stars
can be obtained by using the similar formula to the equation(\ref{ele}),
but with the different coefficient $a_0$. The coefficient can be generally defined by
$M_a=\int d^3x m_a^2a(x)^2/2\propto m_a^2f_a^2a_0^2\times \mbox{``volume"}$, 
where the ``volume" denotes the volume of the axion stars.
Because the volume of the stretched ( spherical ) axion stars is approximately 
given by $0.25\times 10^5$km$^3$ ( $10^6\rm km^3$ ), 
the value $a_0$ for the stretched axion stars is $2\sqrt{10}$ times larger than the value in eq(\ref{a_0}) for
the spherical axion stars.
Thus, the electric fields on the stretched axion stars are $2\sqrt{10}$ times stronger than those on the
spherical axion stars. Although the strengths of the electric fields are different, 
both electric fields gives rise to essentially identical results about   
how fast the axion stars evaporate into radiations. 
Actually,
the emission rate $\dot{W}/(\lambda)^3$ for the stretched axion stars is $40$ times larger than
the value in eq(\ref{w}) for the spherical axion stars. 
On the other hand, the energy density ( $=M_a/\rm ``volume" $ ) of such axion stars 
is $40$ times larger than
the one of the spherical axion stars. Therefore,
the stretched axion stars evaporate in the identical depth with $n_e\sim 10^{20}\rm cm^{-3}$
to the depth in which the spherical axion stars evaporate. 

\vspace{0.1cm}
As we have shown with explicit calculations\cite{i} of f-f absorptions, 
the atmospheres even with dense electrons $n_e\sim 10^{20}\rm cm^{-3}$ 
are transparent for the radio
waves with low frequencies $m_a/2\pi\simeq 2.4\mbox{GHz} (m_a/10^{-5}\rm eV)$.
This is because there are strong magnetic fields $B\sim 10^{10}$G so that
cyclotron energies 
$eB/m_e$ \big( $\sim 10^2\mbox{eV}(B/10^{10}\rm G)$ \big) or 
$eB/M_p$ \big( $\sim 10^{-1}\mbox{eV}(B/10^{10}\rm G$) \big)
are much larger than the energy 
$m_a/2\pi$ ( $ \sim 10^{-5}$eV ) of the radiations.
Thus, the radiations can pass through the atmospheres.

\vspace{0.1cm}
We proceed to discuss how amount of the energies is emitted and
how long the emission lasts.
The radiations are dipole radiations and mainly emitted into the direction
perpendicular to the electric or magnetic fields $\vec{E} \propto \vec{B}$. Thus,
the half of the amount of the radiations can pass
through the atmosphere, while the other half is absorbed by the inside of
the neutron stars. 
These absorbed radiations deposit large amount of the energies $\sim 10^{25}$ GeV/cm$^3$ in
the small regions in an instant such as $10^{-10}$s. Furthermore, 
even if only a small fraction, e.g. $10^{-5}$ of
the radiations passing through the atmospheres are absorbed in the atmosphere,
the atmosphere may be exploded. Then, the emission stops.
For example, 
if the energy $10^{-5}\times 10^{25}\mbox{GeV/cm$^3$}$ is absorbed in the
electrons $10^{20}\rm cm^{-3}$, each electrons gain the energy $1$ GeV
so that the atmosphere is exploded.
Therefore, although it takes $0.1$ second for the whole of the stretched axion stars to fall in
the surface of the neutron star, the emission may
last only for a fraction of $0.1$ second, e.g. $10^{-3}$s.
It would be difficult to estimate how long the emission lasts. But, if the emission
lasts for $10^{-3}$s as observed, the amount of the radiant energies is given by 
$10^{-3}\mbox{s}/0.1 \mbox{s} \times M_a\simeq 10^{40}$erg.
Thus, we can see the consistency between 
the total energies and the durations of the observed FRBs.
In this way we find that almost all of the previous results hold even if we take account
of the effects of the tidal forces.

% It should be mentioned
% that although the radiations are linearly polarized when they are emitted,
% some of them are circularly polarized after
% they pass through the
% magnetospheres of neutron stars.
% The magnetospheres are composed of electrons or positrons, 
% which are produced by the Schwinger mechanism under electric field
% generated by the rotation of the magnetic field $B$.
% The charged particles are distributed to screen the electric field. 
% The spatial distribution of the electrons is different from the distribution of the positrons.
% The number density of electrons ( positrons ) in the magnetospheres
% is given by the Goldreich-Julian density 
% $\simeq \Omega B/2\pi\sim 10^7$cm$^{-3}\big(\Omega/(2\pi/\mbox{s})\big)\big(B(r)/10^{10}\mbox{G}\big)$ 
% with angular velocity $\Omega$ of
% neutron stars.
% These electrons ( positrons ) absorb right ( left ) handed circularly polarized radiations
% when the cyclotron frequency $\omega=eB(r)/m_e$ becomes equal to $m_a/2\pi$, respectively.
% This is because magnetic fields $B(r)$ decreases such that $B(r)\propto 1/r^3$.

\vspace{0.2cm}
Finally, we would like to show why the radiations are not monochromatic, although
it apparently seems that the radiations are monochromatic; their frequencies are
given by the axion mass. 
The FRBs have been observed at the frequencies 
$1.2$GHz$\sim 1.6$GHz. In our model
they are dipole radiations emitted by electrons harmonically oscillating.
These electrons in the neutron stars have temperatures
of the order of $10^5\mbox{K}\simeq 10$ eV or larger. 
Furthermore, the number density of the electrons is very large; $n_e\sim 10^{20}\rm cm^{-3}$.  
Thus, the line spectrum is broadened mainly because of the thermal and collisional effects.
As long as the temperatures are less than $10^{9}$K $\sim 10^5$eV, the thermal
effects are smaller than the collisional effects in the atmospheres with dense
electrons $n_e\sim 10^{20}\rm cm^{-3}$. 

The radiations are emitted by electrons whose
density $n_e$ is large such as $ 10^{20}\rm cm^{-3}$.
These electrons interact with each others. 
The amplitude $x_e$ of the oscillating electrons is approximately given by 
$e\alpha a_0B/m_a^2m_e\pi \sim 0.3$cm ($B/10^{10}$G).
On the other hand, 
the cyclotron radius $l_e$ of the electrons is equal to 
$\sqrt{1/eB}\simeq 0.4\times 10^{-8}$cm $(10^{10}\mbox{G}/B)^{1/2}$. Thus,
we may roughly estimate the scattering cross section among electrons such as $\pi l_e^2\alpha^2$,
with $\alpha\simeq 1/137$. 
Then, the volume $\pi l_e^2\alpha^2x_e$ swepted by the oscillating electrons is equal to 
$0.3\pi\,\rm cm \times (0.4\times 10^{-8}cm)^2\alpha^2\sim 10^{-21}\rm cm^3$. 
The number of electrons in the volume is equal to
$n_e\pi l_e^2x_e\alpha^2\simeq 10^{20}\rm cm^{-3} 
\times 10^{-21}\mbox{cm$^3$}\simeq 10^{-1}(n_e/10^{20}\rm cm^{-3})$. 
Therefore, we find that 
approximately $(2.4\times 10^9\mbox{s}^{-1})\times 10^{-1}(n_e/10^{20}\mbox{cm}^{-3})
\simeq 2.4\times 10^8(n_e/10^{20}\rm cm^{-3})$ 
times collisions take place in a second. 
They cause the radiations to have bandwidth $\sim 0.24$GHz.
The bandwidths depend on the densities of electrons
which emit most of the radiations. 
Although the estimation is very rough, we can see that
the collisions among electrons give rise to finite bandwidths. 
The evidence of the finite bandwidths has recently been
suggested\cite{MWA}; there are no detection of FRBs with low frequencies $\sim 140$MHz.  

\vspace{0.1cm}
We should mention that the observed radiations are redshifted in several ways.
The frequency of the electric fields induced on the axion stars under the magnetic fields
is equal to $\omega=m_a/2\pi\simeq 2.4$GHz$(m_a/10^{-5}\rm eV)$.
Since the axion stars collide with the neutron stars at the relative velocity $\sqrt{2GM_{ns}/R_{ns}}$,
the radiations are redshifted such as
$\omega_{ns}=\omega\sqrt{1-2GM_{ns}/R_{ns}}\simeq 0.76\,\omega$ 
at the rest frame of the neutron stars with $M_{ns}=1.4M_{\odot}$ and $R_a=100$km.
Thus, the radiations with the frequency $\omega_{ns}$
are emitted at the rest frame of the neutron stars.
These radiations are gravitationally redshifted when we observe them
far from the neutron stars so that their frequency is given by 
$\omega'=\omega_{ns}\sqrt{1-2GM_{ns}/R_{ns}}$.
Finally, the frequency of the radiations observed at the earth is
equal to 
$\omega_{ob}=\omega'/(1+z)=\omega(1-2GM_{ns}/R_{ns})/(1+z)\simeq 1.38\mbox{GHz}/(1+z)(m_a/10^{-5}\rm eV)$
when the neutron stars are located at the redshift $z$.

\vspace{0.2cm}

We have discussed the collision between a single neutron star and an axion star.
On the other hand, it is possible for an axion star to collide with a binary neutron star.
Then, the tidal force of the binary neutron star intricately deforms the axion star so that its form
does not simply look like a long stick.
Furthermore, the axion star may collide with both of the neutron stars. The collision
would cause a two-component FRB, which has been recently observed\cite{two}.

Radio bursts similar to the  
FRBs may arise from the collisions between magnetic white dwarfs\cite{wd} and the axion stars.
Some of the white dwarfs have strong magnetic fields $B\sim 10^9$G so that
radio bursts can be produced in the atmospheres of the white dwarfs. But there is a
difference between FRBs from neutron stars and the radio bursts from white dwarfs.
Because gravitational redshifts in the white dwarfs are much smaller than those in the neutron stars,
the frequency $\omega_{ob}\simeq 2.4\mbox{GHz}/(1+z)(m_a/10^{-5}\rm eV)$ of the radiations from the white dwarfs
is higher than the frequency $\omega_{ob}\simeq 1.38\mbox{GHz}/(1+z)(m_a/10^{-5}\rm eV)$
of FRBs from neutron stars.

\vspace{0.2cm}
We have examined our production mechanism of the FRBs by taking account of the tidal forces of the neutron stars.
Although the tidal forces stretch the spherical axion stars to look like long sticks, 
our production mechanism works well as a model of the progenitors.
Because the dark matter concentrates in the centers of galaxies, 
the collisions between dark matter axion stars and
neutron stars would most frequently take place in the centers. It seems that recent observations\cite{center} of the FRBs
show such a possibility of FRBs arising from the centers.

\vspace{0.2cm}
The author
expresses thanks to members of theory center, KEK for their hospitality.

%%%%%%%%%%%%%%%%%%%%%%%

\end{document}